\documentclass[10pt,a4paper]{article}
\usepackage[utf8]{inputenc}
\usepackage[english]{babel}

\usepackage{amsmath}
\usepackage{amsfonts}
\usepackage{amssymb}

\usepackage[colorlinks,citecolor=blue,urlcolor=blue,linkcolor=blue]{hyperref}

\usepackage[left=2cm,right=2cm,top=2cm,bottom=2cm]{geometry}

\usepackage{authblk}

\usepackage{feynmp}
\newcommand{\abs}[1]{\left\lvert #1 \right\rvert}

\title{Scalaron Decay in Perturbative Quantum Gravity}
\author[1,2]{B. Latosh \thanks{  \href{mailto:latosh@theor.jinr.ru}{latosh@theor.jinr.ru}  }  }
\affil[1]{Bogoliubov Laboratory of Theoretical Physics, JINR, Dubna 141980, Russia}
\affil[2]{Dubna State University, Universitetskaya str. 19, Dubna 141982, Russia}
\date{}

\begin{document}

\maketitle

\begin{abstract}
  A certain quadratic gravity model provides a successfully inflationary scenario. The inflation is driven by the new scalar degree of freedom called scalaron. After the end of inflation the scalaron decays in matter and dark matter degrees of freedom reheating the Universe. We study new channels by which the scalaron can transfer energy to the matter sector. These channels are annihilation and decay via intermediate graviton states. Results are obtained within perturbative quantum gravity. In the heavy scalaron limit only scalar particles are produced by the annihilation channel. Scalaron decays in all types of particles are allowed. In the light scalaron limit decay channel is strongly suppressed. Boson production via the annihilation channel is expected to be dominant at the early stages of reheating, while fermion production will dominate later stages.
\end{abstract}

\section{Introduction}

The modified gravity $R+R^2$ model has a special place in the modified gravity landscape. Firstly, the model is healthy despite having higher derivative terms \cite{Accioly:2000nm,Hindawi:1995an}. Secondly, the model has one additional scalar degree of freedom which can be made explicit via a one-to-one mapping on a scalar-tensor gravity \cite{Hindawi:1995an,Dicke:1961gz,Maeda:1988ab,Faraoni:1998qx,DeFelice:2010aj}. Most importantly, the model provides an inflationary scenario completely consistent with the observational data \cite{Starobinsky:1980te,Planck:2018jri,BICEP:2021xfz,Paoletti:2022anb}.

Opportunities to describe the universe reheating within the $R+R^2$ model were extensively studied. Perhaps the simplest scenario is based on transmutations of the new gravitational degree of freedom -- scalaron \cite{Vilenkin:1985md,Koshelev:2016xqb,Arbuzova:2011fu,Arbuzova:2018ydn,Arbuzova:2021etq}. Scalaron drives inflation in the slow roll regime sliding on the plane part of the potential. When the inflation ends with a graceful exit, the scalaron begins to oscillate around the minimum of the potential. These oscillations can induce an extensive particle creation which will reheat the universe. If the scalaron has channels through which it can transmute into other degrees of freedom, then the corresponding factors, like an annihilation cross section or a decay width, will naturally enter cosmological equations and will serve as dumping factors suppressing the discussed oscillations. The exact mechanism responsible for the scalaron transmutation is strongly model-dependent.

We address two aspects of such a reheating scenario. First and foremost, scalaron can annihilate to matter (and dark matter) degrees of freedom through intermediate graviton states. Such annihilation is possible because all types of matter, including the scalaron, are coupled to gravity. Such processes can be consistently described within the perturbative quantum gravity and the corresponding annihilation cross sections can be calculated. The role of this reheating channel will be discussed. Secondly, scalaron can decay directly to matter degrees of freedom because it is universally coupled to the matter energy-momentum tensor. It receives a new decay channel at the one-loop level, as it also can annihilate via intermediate graviton states. We will present particular examples of such processes and discuss their contribution to reheating.

This paper is organized as follows. Firstly, we discuss the setup used in calculations. We show that after the end of an inflation weak quantum gravitational effects can be consistently described by perturbative quantum gravity. We briefly discuss its formalism and applicability in Section \ref{section_setup}. In Section \ref{section_cross_section} we discuss the scalaron annihilation to matter degrees of freedom. For the sake of simplicity, we only consider decays in states with $s=0$, $m\not =0$ which serves as an analogy with the Higgs boson; $s=1/2$, $m\not=0$ which serves as an analogy with quark, lepton, and dark matter degrees of freedom; and $s=1$, $m=0$ which serves as an analogy with gluons and photons. In Section \ref{section_decay} we construct explicit examples of scalaron decays in matter degrees of freedom that only exist at the one-loop level. We discuss the role of such processes in reheating. We bring our conclusions in Section \ref{section_conclusions}.

\section{Perturbative quantum gravity setup}\label{section_setup}

Perturbative quantum gravity provides a simple framework capable to account for quantum gravitational effects consistently \cite{Latosh:2020ysu,Burgess:2003jk,Levi:2018nxp,Calmet:2013hfa,Vanhove:2021zel}. The approach is based on the following premises. Firstly, the constructed theory is effective. This means that the theory applicability domain lies below the Planck scale and it should not be extended further in the UV. Secondly, gravity is perturbative. This means that the theory only accounts for gravitational effects described by small metric perturbations.

On the practical ground these premises results in the following construction. The full spacetime metric $g_{\mu\nu}$ is composite of the flat background $\eta_{\mu\nu}$ and small perturbations $h_{\mu\nu}$:
\begin{align}
  g_{\mu\nu} = \eta_{\mu\nu} + \kappa \, h_{\mu\nu} \,.
\end{align}
Here $\kappa$ is the gravitational coupling related with the Newton constant $G_N$ as follows:
\begin{align}
  \kappa^2 = 32\,\pi\, G_N .
\end{align}
In some sense $h_{\mu\nu}$ plays a role of a spin-$2$ gauge field propagating in a flat background. Quantum dynamics of the system is described by the corresponding generating functional
\begin{align}
  \mathcal{Z}[J^{\mu\nu}] = \int\mathcal{D}\left[h_{\alpha\beta}\right] \,\exp\left[ i\,\mathcal{A}[h_{\alpha\beta}] + i\, J^{\mu\nu} h_{\mu\nu} \right] .
\end{align}
Here $\mathcal{A}$ is the microscopic action describing the used gravity model. The action $\mathcal{A}$ shall be expanded in a perturbative series with respect to $h_{\mu\nu}$ to spawn the following infinite series:
\begin{align}
  \begin{split}
    \mathcal{A} = &-\cfrac12\, h_{\mu\nu} \mathcal{P}^{\mu\nu\alpha\beta} \,\square h_{\alpha\beta} \\
    &+ \kappa \,\widehat{\mathcal{V}}_3^{\mu_1\nu_1\mu_2\nu_2\mu_3\nu_3} h_{\mu_1\nu_1} h_{\mu_2\nu_2} h_{\mu_3\nu_3} + \kappa^2 \,\widehat{\mathcal{V}}_4^{\mu_1\nu_1\mu_2\nu_2\mu_3\nu_3\mu_4\nu_4} h_{\mu_1\nu_1} h_{\mu_2\nu_2} h_{\mu_3\nu_3} h_{\mu_4\nu_4} +\mathcal{O}\left(\kappa^3\right) \,.
  \end{split}
\end{align}
The first term of the expansion describes the graviton propagator, and terms $\widehat{\mathcal{V}}_n$ correspond to graviton interaction vertices. It shall be noted that this expansion contains an infinite number of interaction terms consecutively suppressed by the Planck scale.

The constructed theory is non-renormalizable \cite{Latosh:2020ysu,tHooft:1974toh,Goroff:1985sz}. Within the discussed approach the absence of renormalizability is explained by the finite applicability domain. An application of the standard renormalization procedure for perturbative gravity would require an infinite number of renormalized constants. This only shows that the data on the UV behavior of gravity is missing from the theory, which is already established by the constraint on its applicability domain.

Despite these disadvantages, the theory provides a consistent way to calculate amplitudes in a controllable way. At any order of the perturbation theory a set of relevant interactions can be identified uniquely together with Feynman graphs contributing to a given matrix element. Feynman rules for graviton and matter interactions, which are the core of the theory, can be evaluated analytically \cite{Prinz:2020nru,DeWitt:1967uc,Sannan:1986tz}. In paper \cite{Latosh:2022ydd} an algorithm evaluating Feynman rules for gravity was proposed. It was implemented in FeynGrav package which extends FeynCalc \cite{Latosh:2022ydd,Mertig:1990an,Shtabovenko:2020gxv}. In this paper all calculations are performed with FeynGrav.

The perturbative approach can be used to account for quantum gravitational effects after the end of an inflationary phase. First and foremost, it is safe to assume that all processes involving scalaron, matter, and dark matter degrees of freedom in a post-inflationary universe occur at a time scale much smaller than the scale of the cosmological expansion. To put it otherwise, such processes are well localized both in space and in time, therefore, one can decouple them from the cosmological background and account only for small local metric perturbations describe by the perturbative theory. Secondly, $R+R^2$ gravity admits a unique mapping of a scalar-tensor gravity \cite{Accioly:2000nm,Hindawi:1995an}. This mapping diagonalizes the model Lagrangian and allows one to use the standard formalism free from higher derivative terms.

Consequently, the perturbative approach to quantum gravity provides a consistent controllable way to account for quantum gravitational effects. Its implementations for a post-inflationary universe are not obstructed and can be used to study processes involving gravitational, scalaron, matter, and dark matter degrees of freedom. 

\section{Scalaron annihilation}\label{section_cross_section}

Let us turn to a discussion of scalaron annihilation to matter degrees of freedom. We consider annihilation of a pair of scalarons (which are scalars with mass $M$) to light scalars $s=0$, $0 < m \ll M$; light Dirac fermions $s=1/2$, $0 < m_\text{f} \ll M$; massless vectors $s=1$, $m_\text{v}=0$. These processes are chosen because such degrees of freedom can be associated both with the standard model degrees of freedom and with dark matter degrees of freedom. Namely, the light scalars can be associated with the Higgs boson, fermionic degrees of freedom can be associated with quarks and leptons, and massless vector degrees of freedom can be associated either with photons or with gluons.

In this section, firstly, we calculate all the discussed amplitudes and the corresponding cross sections. After this, we analyze the physical implications of the obtained results. All calculations are done with packages ``FeynCalc'' \cite{Mertig:1990an,Shtabovenko:2020gxv}, ``FeynGrav'' \cite{Latosh:2022ydd}, ``Package-X'' \cite{Patel:2015tea,Patel:2016fam}, and ``FeynHelpers'' \cite{Shtabovenko:2016whf}. The corresponding publications contain detailed descriptions of their usage, so we will not discuss them further.

Kinematics of such annihilation processes is given by the following:
\begin{align}\label{kinematics}
  \begin{split}
    &
    \begin{cases}
      p_1^\mu &=\begin{pmatrix} \sqrt{M^2+p^2} & 0 & 0 & p \end{pmatrix} \,, \\
      p_2^\mu &=\begin{pmatrix} \sqrt{M^2+p^2} & 0 & 0 & -p \end{pmatrix} \,, \\
      q_1^\mu &=\begin{pmatrix} \sqrt{m^2+q^2} & q\,\sin\theta & 0 & q\,\cos\theta \end{pmatrix} \,,\\
      q_2^\mu &=\begin{pmatrix} \sqrt{m^2+q^2} & -q\,\sin\theta & 0 & -q\cos\theta \end{pmatrix} \,,
    \end{cases}\\
    & \hspace{25pt} q=\sqrt{M^2+p^2-m^2}\,\\
  \end{split}
  &
  \begin{cases}
    s &=(p_1+p_2)^2 = 4 \left(M^2+p^2\right) \,, \\
    t &=(p_1-q_1)^2= -\left(p+q\right)^2 + 4 \,p\,q\,\cos^2\frac{\theta}{2}\,, \\
    u &=(p_1-q_2)^2 =-(p+q)^2 + 4 \,p\,q\,\sin^2\frac{\theta}{2}\,.
  \end{cases}
\end{align}
Here $p_1$ and $p_2$ are in-going on-shell momenta of scalarons; $q_1$ and $q_2$ are out-going momenta of produced particles; $p$ is the center-of-mass spacial momentum; $m$ is the mass of produced degrees of freedom; $s$, $t$, and $u$ are the standard Mandelstam variables \cite{Mandelstam:1959bc,Mandelstam:1958xc}.

A given annihilation matrix element $\mathcal{M}$ is related with the differential cross section by the standard formula:
\begin{align}
  d\sigma = \cfrac{1}{j} \,\abs{\mathcal{M}}^2 \,(2\pi)^4 \,\delta^{(4)}(p_1+p_2-q_1-q_2)\,\cfrac{1}{2\,E(p_1)}\,\cfrac{1}{2\, E(p_2)}\,\cfrac{d^3 q_1}{(2\pi)^3\,2\,E(q_1)}\,\cfrac{d^3 q_2}{(2\pi)^3\,2\,E(q_2)}\,.
\end{align}
Here $j$ is the factor normalizing the cross section for a unit flow:
\begin{align}
  \cfrac{1}{j} = \left(\cfrac{\sqrt{p_1\cdot p_2 - M^2\,M^2}}{p_1^0\,p_2^0}\right)^{-1} = \cfrac{M^2+p^2}{\sqrt{\left( M^2 + 2 \,p^2\right)^2 - M^4}} \,.
\end{align}
The other factors normalize free initial and finial states, and perform an integration over the two-body phase space. This formula is discussed in classical textbooks \cite{Bilenky:1995zq,Weinberg:1995mt,Peskin:1995ev} in great details so we will not discuss it further. For the given case the formula provides the following relation between a matrix element and the differential cross section:
\begin{align}
  \cfrac{d\sigma}{d\Omega} = \cfrac{1}{256\,\pi^2}\,\cfrac{1}{M}\,\cfrac{1}{p}\,\cfrac{\sqrt{1 + \left(\frac{p}{M}\right)^2 - \left(\frac{m}{M}\right)^2}}{1+\left(\frac{p}{M}\right)^2} ~\abs{\mathcal{M}}^2.
\end{align}
Here $d\Omega = \sin\theta\,d\theta \,d\varphi$ is the solid angle.

The annihilation of two scalarons in two light scalars with mass $0<m\ll M$ is given by the following matrix element:
\begin{align}\label{amplitude_1}
  \begin{gathered}
    \begin{fmffile}{D1}
      \begin{fmfgraph*}(40,40)
        \fmftop{T1,T2}
        \fmfbottom{B1,B2}
        \fmf{dots}{B1,V1,B2}
        \fmf{dbl_wiggly}{V1,V2}
        \fmf{dashes}{T1,V2,T2}
        \fmfdot{V1,V2}
        \fmflabel{$p_1$}{B1}
        \fmflabel{$p_2$}{B2}
        \fmflabel{$q_1$}{T1}
        \fmflabel{$q_2$}{T2}
      \end{fmfgraph*}
    \end{fmffile}
  \end{gathered}
  =\mathcal{M}_{\text{SS}\to\text{ss}} =  i\, \cfrac{\kappa^2}{4} \,\cfrac{1}{s} \left[ t\,u - (M^2 +m^2 ) (t+u) + m^4 + M^4 + 4\, m^2\, M^2\right] .
\end{align}
Here and below in this section notations \eqref{kinematics} are used. Therefore, $p_1$ and $p_2$ are in-going on-shell scalaron momenta, $q_1$ and $q_2$ are out-going on-shell momenta of scalars, and these momenta are subjected to the conservation law $p_1+p_2 = q_1 + q_2$. The amplitude produces the following differential cross section:
\begin{align}
  \begin{split}
    \cfrac{d\sigma_{\text{SS}\to\text{ss}}}{d\Omega} = \cfrac{1}{16}\,(G M)^2\,\cfrac{M}{p}\, \cfrac{\sqrt{1+\left(\!\frac{p}{M}\!\right)^2-\left(\!\frac{m}{M}\!\right)^2}}{\left[1+\left(\!\frac{p}{M}\!\right)^2\right]^3} \,\Bigg[&\left(1+\left(\!\frac{p}{M}\!\right)^2\right)\left(2+\left(\!\frac{p}{M}\!\right)^2+\left(\!\frac{m}{M}\!\right)^2\right) \\
      &- \left(\!\frac{p}{M}\!\right)^2\left(1+\left(\!\frac{p}{M}\!\right)^2-\left(\!\frac{m}{M}\!\right)^2\right) \cos2\theta\Bigg]^2\,.
  \end{split}
\end{align}
The full cross section reads:
\begin{align}\label{cross_section_1}
  \begin{split}
    \sigma_{\text{SS}\to\text{ss}} =\cfrac{\pi}{60}\,(G M)^2\,\cfrac{M}{p}\,\cfrac{\sqrt{1+\left(\!\frac{p}{M}\!\right)^2-\left(\!\frac{m}{M}\!\right)^2}}{\left[1+\left(\!\frac{p}{M}\!\right)^2\right]^3} \Bigg[ 15\,\left(1+\left(\!\frac{p}{M}\!\right)^2\right)^2\left(2+\left(\!\frac{p}{M}\!\right)^2+\left(\!\frac{m}{M}\!\right)^2\right)^2\\
      + 10\, \left(1+\left(\!\frac{p}{M}\!\right)^2\right)\left(2+\left(\!\frac{p}{M}\!\right)^2+\left(\!\frac{m}{M}\!\right)^2\right) \left(\!\frac{p}{M}\!\right)^2\left(1+\left(\!\frac{p}{M}\!\right)^2-\left(\!\frac{m}{M}\!\right)^2\right)\\
       + 7\, \left(\!\frac{p}{M}\!\right)^4\left(1+\left(\!\frac{p}{M}\!\right)^2-\left(\!\frac{m}{M}\!\right)^2\right)^2 \Bigg]\,.
  \end{split}
\end{align}
It is useful to expand \eqref{cross_section_1} in a series with respect to small mass $m \ll M$ and small momentum $p \ll M$.
\begin{align}
  \begin{split}
    \sigma_{\text{SS}\to\text{ss}} =& \cfrac{\pi}{15}\,(G M)^2\,\cfrac{M}{p} \cfrac{1}{\sqrt{1 +\left(\!\frac{p}{M}\!\right)^2}} \, \left[ 15 + 20\,\left(\!\cfrac{p}{M}\!\right)^2 + 8 \left(\!\cfrac{p}{M}\!\right)^4 \right] + \mathcal{O}\left(\cfrac{m^2}{M^2}\right)\\
    =& \cfrac{\pi}{4} \,(G M)^2\,\cfrac{M}{p}\,\sqrt{1-\left(\!\cfrac{m}{M}\!\right)^2}\,\left[2+\left(\!\cfrac{m}{M}\!\right)^2\right]^2 + \mathcal{O}\left(\cfrac{p}{M}\right) \\
    =& \left[\pi\,(G M)^2\,\cfrac{M}{p} + \mathcal{O}\left(\cfrac{m^2}{M^2}\right)\right] + \mathcal{O}\left(\cfrac{p}{M}\right)\,.
  \end{split}
\end{align}
We will discuss this result in detail below at the end of this section. For the time begin we only highlight two features of cross section \eqref{cross_section_1}. Firstly, the cross section has a smooth $m\to 0$ limit, so it can be used for massless scalars. Secondly, the cross section is expectedly singular in $p\to 0$ limit. Both matrix element \eqref{amplitude_1} and cross section \eqref{cross_section_1} are evaluated in the center-of-mass frame. Limit $p\to 0$ corresponds to a situation when particles do not approach each other and do not participate in the interaction, therefore the corresponding matrix element is expected to be singular.

The annihilation of two scalarons in two light Dirac fermions of mass $0< m_\text{f} \ll M$ is given by the following matrix element:
\begin{align}\label{amplitude_2}
  \begin{gathered}
    \begin{fmffile}{D2}
      \begin{fmfgraph*}(40,40)
        \fmftop{T1,T2}
        \fmfbottom{B1,B2}
        \fmf{dots}{B1,V1,B2}
        \fmf{dbl_wiggly}{V1,V2}
        \fmf{fermion}{T2,V2,T1}
        \fmfdot{V1,V2}
        \fmflabel{$p_1$}{B1}
        \fmflabel{$p_2$}{B2}
        \fmflabel{$q_1$}{T1}
        \fmflabel{$q_2$}{T2}
      \end{fmfgraph*}
    \end{fmffile}
  \end{gathered}
  =\mathcal{M}_{\text{SS}\to\text{f}\overline{\text{f}}} = u(q_1) \left[i\, \cfrac{\kappa^2}{16}\,\cfrac{1}{s}\, (s+2\,M^2)(\hat{q}_1+\hat{q}_2+ 4\,m_\text{f}) \right] v(q_2) .
\end{align}
Here the same notations for momenta are used and $u(q)$, $v(q)$ are spinors describing polarization of external fermion states. The corresponding differential cross section reads:
\begin{align}
  \cfrac{d\sigma_{\text{SS}\to\text{f}\overline{\text{f}}}}{d\Omega}=\cfrac12\,(G M)^2\, \left(\cfrac{m_\text{f}}{M}\right)^2 \cfrac{M}{p} ~ \cfrac{\left[3 + 2 \left(\!\frac{p}{M}\!\right)^2\right]^2 \,\left[1+\left(\!\frac{p}{M}\!\right)^2 - \left(\!\frac{m_\text{f}}{M}\!\right)^2\right]^{\frac32}}{\left[1+\left(\!\frac{p}{M}\!\right)^2\right]^3} \,.
\end{align}
Here a summation over all external fermion polarizations is performed. The complete cross section is given by the following formula:
\begin{align}\label{cross_section_2}
  \sigma_{\text{SS}\to\text{f}\overline{\text{f}}}= 2\,\pi\,(G M)^2\, \left(\cfrac{m_\text{f}}{M}\right)^2\,\cfrac{M}{p} \cfrac{\left[3 + 2 \left(\!\frac{p}{M}\!\right)^2\right]^2 \,\left[1+\left(\!\frac{p}{M}\!\right)^2 - \left(\!\frac{m_\text{f}}{M}\!\right)^2\right]^{\frac32}}{\left[1+\left(\!\frac{p}{M}\!\right)^2\right]^3} \,.
\end{align}
Leading order contributions in small mass $m_\text{f} \ll M$ and small momentum $p\ll M$ read:
\begin{align}
  \begin{split}
    \sigma_{\text{SS}\to\text{f}\overline{\text{f}}}=& 2\,\pi\,(G M)^2\,\cfrac{M}{p}\,\left(\cfrac{m_\text{f}}{M}\right)^2\, \cfrac{\left[3+2\,\left(\!\frac{p}{M}\!\right)^2\right]^2}{\left[1+\left(\!\frac{p}{M}\!\right)^2\right]^{\frac32}} + \mathcal{O}\left( \cfrac{m_\text{f}^4}{M^4}\right) \\
    =&18\,\pi\,(G M)^2\,\cfrac{M}{p}\,\left(\cfrac{m_\text{f}}{M}\right)^2\,\left[1-\left(\cfrac{m_\text{f}}{M}\right)^2\right]^{\frac32} + \mathcal{O}\left(\cfrac{p}{M}\right) \\
    =& \left[ 18\,\pi\,(G\,M)^2\,\cfrac{M}{p}\,\left(\cfrac{m_\text{f}}{M}\right)^2 + \mathcal{O}\left(\cfrac{p}{M}\right)\right] + \mathcal{O}\left(\cfrac{m_\text{f}^4}{M^4}\right) \,.
  \end{split}
\end{align}
In full analogy with the previous case, the cross section is singular in $p\to 0$ limit and admits a smooth $m_\text{f} \to 0$ limit. In a contrast with the previous case, the cross section vanishes if $m_\text{f} =0$, thus massless fermions are not produced via this mechanism.

Finally, the amplitude describing an annihilation of two scalarons in a pair of massless vectors reads:
\begin{align}\label{amplitude_3}
  \begin{gathered}
    \begin{fmffile}{D3}
      \begin{fmfgraph*}(40,40)
        \fmftop{T1,T2}
        \fmfbottom{B1,B2}
        \fmf{dots}{B1,V1,B2}
        \fmf{dbl_wiggly}{V1,V2}
        \fmf{photon}{T1,V2,T2}
        \fmfdot{V1,V2}
        \fmflabel{$p_1$}{B1}
        \fmflabel{$p_2$}{B2}
        \fmflabel{$q_1$}{T1}
        \fmflabel{$q_2$}{T2}
      \end{fmfgraph*}
    \end{fmffile}
  \end{gathered} =\mathcal{M}_{\text{SS}\to\text{v}\overline{\text{v}}} \,,
\end{align}

\begin{align}
  \begin{split}
    \mathcal{M}_{\text{SS}\to\text{v}\overline{\text{v}}} = \left(i\,\cfrac{\kappa^2}{4}\,\cfrac{1}{s}\right) \varepsilon_{\sigma_1}^*(q_1)\,\varepsilon_{\sigma_2}^*(q_2) \,\Bigg[  M^4 \eta^{\sigma_1\sigma_2} + M^2 \Big\{ (s-t-u) \eta^{\sigma_1\sigma_2} -q_1^{\sigma_1}q_1^{\sigma_2}-q_2^{\sigma_1}q_2^{\sigma_2} - 4\,q_1^{\sigma_2} q_2^{\sigma_1} \Big\}\\
      -\cfrac12\, (s^2 - t^2 - u^2)\, \eta^{\sigma_1\sigma_2} + s \left\{ p_1^{\sigma_1} p_2^{\sigma_2} + p_1^{\sigma_2} p_2^{\sigma_1} + q_1^{\sigma_2} q_2^{\sigma_1} \right\} + t \left(p_1^{\sigma_1} q_1^{\sigma_2} + p_2^{\sigma_2} q_2^{\sigma_1}\right) + u \left( p_1^{\sigma_2} q_2^{\sigma_1} + p_2^{\sigma_1} q_1^{\sigma_2}\right) \Bigg].
  \end{split}
\end{align}
Kinematics \eqref{kinematics} admits a smooth $m\to 0$ limit, so we apply it for this case and use the same notations. Here $\varepsilon_\sigma(q)$ are polarization vectors describing polarizations of free vector states. The corresponding differential cross section reads:
\begin{align}
  \cfrac{d\sigma_{\text{SS}\to\text{v}\overline{\text{v}}}}{d\Omega} = \cfrac12\, (G M)^2\,\cfrac{\left(\!\frac{p}{M}\!\right)^3}{\sqrt{1+\left(\!\frac{p}{M}\!\right)^2}}\, \sin^4\theta\,.
\end{align}
A summation over all external vector polarizations is performed. The full cross section is given by the following formula:
\begin{align}\label{cross_section_3}
  \sigma_{\text{SS}\to\text{v}\overline{\text{v}}} = \cfrac{16}{15}\,\pi \, (G M)^2 \,\cfrac{\left(\!\frac{p}{M}\!\right)^3}{\sqrt{1+\left(\!\frac{p}{M}\!\right)^2}} = \cfrac{16}{15}\,\pi \, (G\,M)^2\,\left(\cfrac{p}{M}\right)^3 + \mathcal{O}\left(\cfrac{p^5}{M^5}\right) \,.
\end{align}
The cross section \eqref{cross_section_3} does not vanish for massless vector bosons similarly to the case of massless scalar bosons. At the same time, the cross section is regular in $p \to 0$ limit and vanishes. 

Let us discuss physical implications of the obtained cross sections \eqref{cross_section_1}, \eqref{cross_section_2}, and \eqref{cross_section_3}. Results are strongly dependent on the mass hierarchy. The obtained cross sections do depend on the mass hierarchy, but its influence goes beyond that. The total energy of the inflaton field at the end of inflation is finite and its value can be constrained by data on the contemporary energy content of the Universe. Consequently, if the inflaton is heavy so its mass lies in the Planck region, then at the end of inflation the universe will be filled by few non-relativistic inflaton particles. On the contrary, if the inflaton is light so its mass lies about the top quark mass, then at the end of inflation the universe will be filled with plenty of relativistic scalarons. In the case of a heavy inflaton mass factors $m/M$, $m_\text{f}/M$ together with the momentum factor $p/M$ are small. In the case of a light inflaton mass factors $m/M$, $m_\text{f}/M$ can be of the order unity while $p/M$ can be large. These cases allow one to draw comprehensive conclusions on the role of such processes in reheating.

Let us start with the case of a heavy scalaron. In than case the leading contributions to the discussed annihilation cross sections read:
\begin{align}
  \begin{split}
    \sigma_{\text{SS}\to\text{ss}} & = \cfrac{1}{16} \, \sigma_0 \, \cfrac{M}{p}  \, , \\
    \sigma_{\text{SS}\to\text{f}\overline{\text{f}}} & = 0 \,, \\
    \sigma_{\text{SS}\to\text{v}\overline{\text{v}}} & = 0 \,.
  \end{split}
\end{align}
Here
\begin{align}
  \sigma_0 \overset{\text{def}}{=}  \pi \, (4\,G\,M)^2\,
\end{align}
is the characteristic cross section equal to a surface of a plain disc with a radius $ 4 \, G\, M$ which is the gravitational radius of a black hole of a mass $2 \, M$. As it was noted above, the production of fermions is strongly suppressed by the factor $\left(m_\text{f}/M\right)^2$ and effectively vanishes no matter the momentum factor $p/M$. The production of massless vectors is free from mass factors, but it is suppressed by the momentum factor $\left( p/M \right)^3$, therefore it also vanishes in the heavy scalaron limit. Therefore, in the heavy scalaron limit only scalar particles are produced effectively.

In the light scalaron limit the leading contributions are given by the following expressions.
\begin{align}
  \begin{split}
    \sigma_{\text{SS}\to\text{ss}} & = \cfrac{1}{30} \,\sigma_0\,\left(\cfrac{p}{M}\right)^2  \, , \\
    \sigma_{\text{SS}\to\text{f}\overline{\text{f}}} & = \cfrac12\,\sigma_0 \,, \\
    \sigma_{\text{SS}\to\text{v}\overline{\text{v}}} & = \cfrac{1}{15}\,\sigma_0 \, \left(\cfrac{p}{M}\right)^2 \,.
  \end{split}
\end{align}
In that case, all cross sections are non-vanishing, but they experience a very different behavior. Production of fermions reaches a certain limit value and does not grow any further. On the contrary, the production of bosons grows quadratic with the center-of-mass momentum $p$. Consequently, in the light scalaron limit production of fermions does not vanish, but it is still suppressed, while scalars and vector degrees of freedom are produced in equal amounts. Let us note that the vector production cross section is bigger than the scalar production cross section. This is because the cross section accounts for two polarizations of vector bosons. Because of this, we prefer to say those bosonic degrees of freedom are produced in equal amounts.

It shall be noted that one does not simply transfer these conclusions for a reheating scenario because of the following. On general grounds, the density of scalarons $n$ is subjected to the Zeldovich-Lee-Weinberg equation \cite{Zeldovich:1965gev,Lee:1977ua}:
\begin{align}
  \dot{n} + 3\, H\, n + \langle \sigma_\text{annihilation} \, v \rangle \left(n^2 - n_0^2\right) =0\,.
\end{align}
Here $H$ is the Hubble parameter, $v=p/M$ is the center-of-mass velocity, $n_0$ is the equilibrium density of the scalarons, and $\langle \cdot \rangle$ notes the thermal average (see \cite{Arbuzova:2021etq} for a recent review). In a given reheating scenario $\sigma_\text{annihilation}$ shall account for all annihilation channels. Our results allows one to obtain explicit expressions for the corresponding thermal average factor. In the heavy scalaron approximation the corresponding factor for bosons reads
\begin{align}
  \left\langle \sigma_{\text{SS}\to\text{ss}}\,v \right\rangle= \left[ \int\limits_0^\infty \,dp\, 4\pi\,p^2\,\exp\left[-\cfrac{p^2}{2\,M\,T} \right]\right]^{-1} \left[\int\limits_0^\infty \,dp\left(\cfrac{1}{16} \,\sigma_0\right) 4\pi\,p^2\,\exp\left[-\cfrac{p^2}{2\,M\,T} \right]\right] =\cfrac{1}{16}\,\sigma_0\,.
\end{align}
In the light scalaron approximation the corresponding factor for scalars (and vector bosons of a given chirality) reads:
\begin{align}
  \left\langle \sigma_{\text{SS}\to\text{ss}}\,v \right\rangle = \left[\int\limits_0^\infty dp\,\cfrac{4\pi\,p^2}{\exp\left[\frac{p}{T}\right]+1}\right]^{-1} \left[\int\limits_0^\infty dp\,\left(\cfrac{1}{30}\,\sigma_0\,\left(\cfrac{p}{M}\right)^3\right)\,\cfrac{4\pi\,p^2}{\exp\left[\frac{p}{T}\right]+1}\right] = \cfrac{31\,\pi^6}{11340\,\zeta(3)}\,\sigma_0\,\left(\cfrac{T}{M}\right)^3\,.
\end{align}
Here $\zeta$ notes the Riemann zeta function. For fermion the factor reads:
\begin{align}
  \left\langle \sigma_{\text{SS}\to\text{f}\overline{\text{f}}}\,v \right\rangle = \left[\int\limits_0^\infty dp\,\cfrac{4\pi\,p^2}{\exp\left[\frac{p}{T}\right]+1}\right]^{-1} \left[\int\limits_0^\infty dp\,\left(\cfrac{1}{2}\,\sigma_0\,\left(\cfrac{p}{M}\right)\right)\,\cfrac{4\pi\,p^2}{\exp\left[\frac{p}{T}\right]+1}\right] = \cfrac{7\,\pi^4}{360\,\zeta(3)} \, \sigma_0 \, \cfrac{T}{M}\,.
\end{align}
This results shows that fermion production may become dominant as it experience a weaker suppression by the thermal factor $(T/M)$.

In concluding this section we summarize the result as follows. The discussed processes and their contribution to reheating are extremely sensitive to the scalaron-matter mass hierarchy. In the heavy scalaron limit production of scalar bosons is dominating. In the light scalaron limit the situation is much more sophisticated. Vector bosons of a given chirality and scalar bosons are produced in equal amounts. Fermions are also produced and it appears that their production may become dominant as they experience much weaker suppression by the thermal factor. A more detailed discussion of a reheating scenario accounting for the presented processes lies beyond the scope of this paper and will be discussed elsewhere.

\section{Scalaron decay}\label{section_decay}

Let us turn to a discussion of scalaron decays that take place only at the one-loop level. The existence of such decay is due to scalaron self-interaction. It is well known that the quadratic gravity
\begin{align}
  \mathcal{A}_\text{quadratic} = \int d^4 x \sqrt{-g} \left(-\cfrac{2}{\kappa^2}\right)\left[R - \cfrac{1}{6\,M^2} \,R^2\right]
\end{align}
can be mapped on a scalar-tensor gravity \cite{Accioly:2000nm,Hindawi:1995an}:
\begin{align}
  \mathcal{A} = \int d^4 x \sqrt{-g} \left[-\cfrac{2}{\kappa^2}\,R + \cfrac12\,(\nabla\phi)^2 - \cfrac{3\,M^2}{\kappa^2}\,\left(\exp\left[\cfrac{\kappa\,\phi}{\sqrt{6}}\right]-1\right)^2\right].
\end{align}
We shall note the in these formulas we omitted the matter energy-momentum tensor which receives a non-minimal coupling to $\phi$ because of the conformal transformations. This coupling is irrelevant to the present problem so we will not discuss it further.

This representation diagonalizes the Lagrangian of the scalar degree of freedom and makes the structure of its self-interaction explicit:
\begin{align}
  \mathcal{A}=\int d^4 x \sqrt{-g} \left[-\cfrac{2}{\kappa^2}\,R + \cfrac12\,(\nabla\phi)^2- \cfrac{M^2}{2}\,\phi^2 -\cfrac{1}{3!} \,\sqrt{\frac{3}{2}} \,\kappa\,M^2 \,\phi^3 + \mathcal{O}\left(\phi^4\right) \right].
\end{align}
This formula shows that the scalaron is a scalar particle with mass $M$ and it admits an infinite number of interaction terms. Each consequent interaction term is suppressed by a higher power of $\kappa$ which makes only the cubic interaction relevant in the low energy limit:
\begin{align}
  \begin{gathered}
    \begin{fmffile}{V3}
      \begin{fmfgraph}(30,30)
        \fmfbottom{B}
        \fmftop{T1,T2}
        \fmf{dots}{T1,V}
        \fmf{dots}{T2,V}
        \fmf{dots}{B,V}
        \fmfdot{V}
      \end{fmfgraph}
    \end{fmffile}
  \end{gathered}
  =-i\, \sqrt{\cfrac{3}{2}}\,\kappa\,M^2\,.
\end{align}
The existence of such a vertex results in an existence of the following processes:
\begin{align}
  \nonumber \\
  \begin{gathered}
    \begin{fmffile}{D7}
      \begin{fmfgraph*}(40,60)
        \fmfbottom{B}
        \fmftop{T1,T2}
        \fmf{dots,tension=2}{B,VB}
        \fmf{dots,right=1,tension=.8}{VB,VU,VB}
        \fmf{dbl_wiggly,tension=2}{VU,VT}
        \fmf{dbl_plain,tension=2}{T1,VT,T2}
        \fmfdot{VB,VU,VT}
        \fmflabel{$p$}{B}
        \fmflabel{$q_1$}{T1}
        \fmflabel{$q_2$}{T2}
      \end{fmfgraph*}
    \end{fmffile}
  \end{gathered}\\ \nonumber
\end{align}
Here the double plain line corresponds to yet unspecified matter degree of freedom. Similar to the previous case we will only consider decays in light scalars, light Dirac fermions, and massless vectors because of the same reasons. Namely, such processes can be easily associated both with the standard model and beyond the standard model degrees of freedom.

The kinematic of a scalaron decay is much more simple. In the center-of-mass frame it is given by the following relations of momenta:
\begin{align}
  \begin{cases}
    p^\mu = \begin{pmatrix} M & 0 & 0 & 0 \end{pmatrix} \,,\\
    q_1^\mu = \begin{pmatrix} \sqrt{m^2+ q^2} & 0 & 0 & q \end{pmatrix} \, ,\\
    q_2^\mu = \begin{pmatrix} \sqrt{m^2+ q^2} & 0 & 0 &-q \end{pmatrix} \,.
  \end{cases}
   & & q= \sqrt{\cfrac{M^2}{4} - m^2}\,.
\end{align}
Here $p$ is the in-going on-shell scalaron momentum, $q_1$ and $q_2$ are out-going on-shell momenta of produced degrees of freedom with mass $m$.

The decay width $\Gamma$ is related with a given matrix element $\mathcal{M}$ by the following formula:
\begin{align}
  d\Gamma =\,\abs{\mathcal{M}}^2\, (2\pi)^4 \delta^{(4)}(p - q_1-q_2) \, \cfrac{1}{2\,M}\,\cfrac{d^3 q_1}{(2\pi)^3\,2\,E(q_1)}\,\cfrac{d^3 q_2}{(2\pi)^3\,2\,E(q_2)}  = \cfrac{1}{64\,\pi^2}\,\cfrac{1}{M}\,\sqrt{1-4\,\left(\cfrac{m}{M}\right)^2} \,\abs{\mathcal{M}}^2\, d\Omega\,.
\end{align}
Derivation of this formula is similar to a derivation of the formula for a differential cross section and discussed in the classical textbooks \cite{Bilenky:1995zq,Weinberg:1995mt,Peskin:1995ev}.

The decay of a scalaron in two light scalars is given by the following matrix element:
\begin{align}
  \nonumber \\
  \begin{gathered}
    \begin{fmffile}{D4}
      \begin{fmfgraph*}(40,60)
        \fmfbottom{B}
        \fmftop{T1,T2}
        \fmf{dots,tension=2}{B,VB}
        \fmf{dots,right=1,tension=.8}{VB,VU,VB}
        \fmf{dbl_wiggly,tension=2}{VU,VT}
        \fmf{dashes,tension=2}{T1,VT,T2}
        \fmfdot{VB,VU,VT}
        \fmflabel{$p$}{B}
        \fmflabel{$q_1$}{T1}
        \fmflabel{$q_2$}{T2}
      \end{fmfgraph*}
    \end{fmffile}
  \end{gathered}
  =\mathcal{M}_{\text{S} \to\text{ss}} \,. \\ \nonumber
\end{align}
Here and below $p$ is the in-going on-shell scalaron momentum; $q_1$ and $q_2$ are out-going on-shell momenta of the scalar field with mass $m$; momenta are connected by the conservation law $p = q_1+q_2$. The amplitude can be calculated in terms of the Passarino-Veltman integrals \cite{Passarino:1978jh,Mertig:1990an}:
\begin{align}
  \begin{split}
    \mathcal{M}_{\text{S} \to\text{ss}} =& \cfrac{\kappa^2}{16}\,\sqrt{\cfrac{3}{2}}\,\kappa\,M^2\, \cfrac{(d-3)(d-2)}{d-1}\, \left( D - 2 + 4 \cfrac{m^2}{M^2} \right) \, i \,\pi^2\,A_0(M^2)\\
    &- \cfrac{\kappa^2}{32}\,\sqrt{\cfrac{3}{2}}\,\kappa\,M^2\, \cfrac{(d-3)(d+2)}{d-1}\,M^2\, \left( D - 2 + 4 \cfrac{m^2}{M^2} \right) \,i\,\pi^2\,B_0(M^2,M^2,M^2)
  \end{split}
\end{align}
The structure of these integrals is well-known and can be evaluated with ``Package-X'' \cite{Patel:2015tea,Patel:2016fam} and ``FeynHelpers'' \cite{Shtabovenko:2016whf} packages for ``FeynCalc'':
\begin{align}
  \mathcal{M}_{\text{S} \to\text{ss}} =-i\,\cfrac{\pi^2\,\kappa^2}{24}\,\sqrt{\cfrac{3}{2}}\,\kappa\,M^2\,(M^2+2\,m^2)\,\left[\cfrac{1}{\varepsilon_\text{UV}}-\ln\cfrac{M^2}{\mu^2}+\cfrac{8}{3} - \gamma - \sqrt{3}\,\pi-\ln\pi\right] -i\, \cfrac{\pi^2\,\kappa^2}{12}\,\sqrt{\cfrac{3}{2}}\,\kappa\,M^2\,m^2 .
\end{align}
Here $\varepsilon_\text{UV}$ is the dim-reg regularization parameter and $\mu$ is the normalization scale.

The amplitude consists of two terms. The first term contains a UV divergence and shall be renormalized. The second term is free from UV divergencies so it shall not be renormalized. It provides a finite contribution to the decay width that is independent of the UV structure of the theory. We will use the following renormalized amplitude:
\begin{align}
  \mathcal{M}_{\text{S} \to\text{ss}} \to \mathcal{M}_{\text{S} \to\text{ss},\text{ren}} = -i\,\cfrac{\pi^2\,\kappa^2}{24}\,\sqrt{\cfrac{3}{2}}\,\kappa\,M^2\,\left(M^2 + 2 m^2\right)\, \mathcal{F}_1 - i\,\cfrac{\pi^2\,\kappa^2}{12}\,\sqrt{\cfrac{3}{2}}\,\kappa\,M^2\,m^2 .
\end{align}
Here $\mathcal{F}_1$ is a finite unknown constant that shall be recovered from empirical data. The corresponding decay width reads:
\begin{align}\label{decay_width_1}
  \Gamma_{\text{S} \to\text{ss}} =& \cfrac{16}{3}\,\pi^6\,\left(G\,M^2\right)^3 \, M \, \sqrt{1-4\left(\cfrac{m}{M}\right)^2} \left[\mathcal{F}_1+2\,(1+\mathcal{F}_1) \left(\cfrac{m}{M}\right)^2\right]^2\,.
\end{align}
Its UV finite part reads:
\begin{align}\label{decay_width_1_UV_finite}
  \Gamma_{\text{S} \to\text{ss}}{}\Big|_\text{UV finite} = \cfrac{64}{3}\,\pi^6\,\left(G\,M^2\right)^3\,M \,\sqrt{1-4\,\left(\cfrac{m}{M}\right)^2}\,\left(\cfrac{m}{M}\right)^4\,.
\end{align}

The decay of a scalar in two light Dirac fermions is described by the following amplitude
\begin{align}
  \nonumber \\
  \begin{gathered}
    \begin{fmffile}{D5}
      \begin{fmfgraph*}(40,60)
        \fmfbottom{B}
        \fmftop{T1,T2}
        \fmf{dots,tension=2}{B,VB}
        \fmf{dots,right=1,tension=.8}{VB,VU,VB}
        \fmf{dbl_wiggly,tension=2}{VU,VT}
        \fmf{fermion,tension=2}{T2,VT,T1}
        \fmfdot{VB,VU,VT}
        \fmflabel{$p$}{B}
        \fmflabel{$q_1$}{T1}
        \fmflabel{$q_2$}{T2}
      \end{fmfgraph*}
    \end{fmffile}
  \end{gathered} = \mathcal{M}_{\text{S}\to\text{f}\overline{\text{f}}}\,. \\ \nonumber
\end{align}
Here the same notations for momenta are used. In terms of the Passarino-Veltman integrals the amplitude reads:
\begin{align}
  \begin{split}
    \mathcal{M}_{\text{S}\to\text{f}\overline{\text{f}}}=u(q_1)\Bigg[&\cfrac{\kappa^2}{16}\,\sqrt{\cfrac32}\,\kappa\,M^2\,(d-2)\,\cfrac{1}{M^2}  \left[(d-3)(\widehat{q}_1+\widehat{q}_2) +2\,(d-2)\,m_\text{f}\right] \,i\,\pi^2\,A_0(M^2)\\
      & -\cfrac{\kappa^2}{32}\,\sqrt{\cfrac32}\,\kappa\,M^2\,(d+2)\,\left[(d-3)(\widehat{q}_1+\widehat{q}_2) +2\,(d-2)\,m_\text{f}\right] \,i\,\pi^2\,B_0(M^2,M^2,M^2) \Bigg] v(q_2)\,.
  \end{split}
\end{align}
Here $u$ and $v$ are spinors describing the polarization of external fermions. The corresponding analytic expression for the amplitude reads:
\begin{align}
  \begin{split}
    \mathcal{M}_{\text{S}\to\text{f}\overline{\text{f}}}= u(q_1)\Bigg[&-i\,\cfrac{\pi^2\,\kappa^2}{16}\,\sqrt{\cfrac32}\,\kappa\,M^2\, \left[\widehat{q}_1+\widehat{q}_2+4\, m_f\right]\left(\cfrac{1}{\varepsilon_\text{UV}} -\ln\cfrac{M^2}{\mu^2} - \sqrt{3}\,\pi -\gamma +3 - \ln\pi\right)\\
    &-i\,\cfrac{\pi^2\,\kappa^2}{4}\,\sqrt{\cfrac32}\,\kappa\,M^2\,m_\text{f}\Bigg] \,v(q_2)\,.
  \end{split}
\end{align}
In full analogy with the previous case the amplitude contains a UV finite part that is not affected by a renormalization. We use the following renormalized expression for the amplitude:
\begin{align}
  \mathcal{M}_{\text{S}\to\text{f}\overline{\text{f}}} \to \mathcal{M}_{\text{S}\to\text{f}\overline{\text{f}},\text{ren}} = u(q_1)\Bigg[-i\,\cfrac{\pi^2\,\kappa^2}{16}\,\sqrt{\cfrac32}\,\kappa\,M^2\,\left[\widehat{q}_1 + \widehat{q}_2 + 4\, m_\text{f}\right]\,\mathcal{F}_2 - i\,\cfrac{\pi^2\,\kappa^2}{4}\,\sqrt{\cfrac32}\,\kappa\,M^2\,m_\text{f} \Bigg] v(q_2)\,.
\end{align}
It provides the following expression for the decay width (a summation over all external fermion polarizations is performed):
\begin{align}\label{decay_width_2}
  \Gamma_{\text{S}\to\text{f}\overline{\text{f}}} = 384\,\pi^6\,\left(G\,M^2\right)^3 \, M \,\left(\cfrac{m_\text{f}}{M}\right)^2 \,\left[1 - 4 \left(\cfrac{m_\text{f}}{M}\right)^2\right]^{\frac32} \, \left(1+\mathcal{F}_2\right)^2 \,.
\end{align}
The part of this decay width that does not depend on the UV structure of the theory reads:
\begin{align}\label{decay_width_2_UV_finite}
  \Gamma_{\text{S}\to\text{f}\overline{\text{f}}}\Big|_\text{UV finite} = 384\,\pi^6\,\left(G\,M^2\right)^3 \, M \,\left(\cfrac{m_\text{f}}{M}\right)^2 \,\left[1 - 4 \left(\frac{m_\text{f}}{M}\right)^2\right]^{\frac32} \,.
\end{align}

Finally, the decay of a scalaron in a pair of massless vectors is given by the following matrix element:
\begin{align}
  \nonumber \\
  \begin{gathered}
    \begin{fmffile}{D6}
      \begin{fmfgraph*}(40,60)
        \fmfbottom{B}
        \fmftop{T1,T2}
        \fmf{dots,tension=2}{B,VB}
        \fmf{dots,right=1,tension=.8}{VB,VU,VB}
        \fmf{dbl_wiggly,tension=2}{VU,VT}
        \fmf{photon,tension=2}{T1,VT,T2}
        \fmfdot{VB,VU,VT}
        \fmflabel{$p$}{B}
        \fmflabel{$q_1$,$\sigma_1$}{T1}
        \fmflabel{$q_2$,$\sigma_2$}{T2}
      \end{fmfgraph*}
    \end{fmffile}
  \end{gathered} = \mathcal{M}_{\text{S}\to\text{v}\overline{\text{v}}} \,. \\ \nonumber
\end{align}
In terms of the Passarino-Veltman integrals the amplitude is given by the following expression:
\begin{align}
  \begin{split}
    \mathcal{M}_{\text{S}\to\text{v}\overline{\text{v}}} =& \varepsilon_{\sigma_1}^* (q_1) \,\varepsilon_{\sigma_2}^* (q_2)\,\cfrac{\kappa^2}{16}\,\sqrt{\cfrac32} \,\kappa\,M^2 \,\left[(d-3)(d-4)\,M^2\,\eta^{\sigma_1\sigma_2} -2 \, (d-3)(d-4)\,q_1^{\sigma_2}q_2^{\sigma_1} - 4 \, q_1^{\sigma_1} q_2^{\sigma_2}\right]\\
    &\times \left[ -\cfrac{d-2}{d-1}\,\cfrac{1}{M^2}\,i\,\pi^2\,A_0(M^2) + \cfrac12\,\cfrac{d+2}{d-1}\,i\,\pi^2\,B_0(M^2,M^2,M^2) \right].
  \end{split}
\end{align}
The corresponding analytic expression reads:
\begin{align}
  \begin{split}
    \mathcal{M}_{\text{S}\to\text{v}\overline{\text{v}}}=\varepsilon_{\sigma_1}^* (q_1) \,\varepsilon_{\sigma_2}^* (q_2)\,\Bigg[& -i\,\cfrac{\pi^2\,\kappa^2}{12}\,\sqrt{\cfrac32}\,\kappa\,M^2 \,q_1^{\sigma_1}\,q_2^{\sigma_2} \left[\cfrac{1}{\varepsilon_\text{UV}} -\ln\cfrac{M^2}{\mu^2} + \cfrac{17}{3} - \gamma-\sqrt{3}\,\pi-\ln\pi\right]\\
      &- i\,\cfrac{\pi^2\,\kappa^2}{24}\,\sqrt{\cfrac32}\,\kappa\,M^2\,M^2\,\eta^{\sigma_1\sigma_2} + i\,\cfrac{\pi^2\,\kappa^2}{12}\,\sqrt{\cfrac32}\,\kappa\,M^2\,q_1^{\sigma_2}\,q_2^{\sigma_1} \Bigg]\,.
  \end{split}
\end{align}
We use the following regularized amplitude
\begin{align}
  \begin{split}
    \mathcal{M}_{\text{S}\to\text{v}\overline{\text{v}}} \to \to \mathcal{M}_{\text{S}\to\text{v}\overline{\text{v}},\text{ren}} = \varepsilon_{\sigma_1}^* (q_1) \,\varepsilon_{\sigma_2}^* (q_2)\,\Bigg[& - i\,\cfrac{\pi^2\,\kappa^2}{12}\,\sqrt{\cfrac32}\,\kappa\,M^2 \,q_1^{\sigma_1}\,q_2^{\sigma_2} \,\mathcal{F}_3 \\
      &-i\, \cfrac{\pi^2\,\kappa^2}{24}\,\sqrt{\cfrac32}\,\kappa\,M^2\,M^2\,\eta^{\sigma_1\sigma_2} + i\,\cfrac{\pi^2\,\kappa^2}{12}\,\sqrt{\cfrac32}\,\kappa\,M^2\,q_1^{\sigma_2}\,q_2^{\sigma_1} \Bigg]\,.
  \end{split}
\end{align}
It produces the following decay width:
\begin{align}\label{decay_width_3}
  \Gamma_{\text{S}\to\text{v}\overline{\text{v}}} = \cfrac{32}{3}\,\pi^6\,(G\,M^2)^3\,M\,.
\end{align}

Unlike the previous cases, the given decay width does not depend on the renormalization constant $\mathcal{F}_3$. The reason behind this is the momentum structure of the amplitude. Vector polarization operators $\varepsilon$ are transverse:
\begin{align}
  \varepsilon_\sigma (q) \,q^\mu =0\,.
\end{align}
Consequently, the part of amplitude that depends on the renormalization vanishes. It should be noted that this feature does only take place on-shell.

Let us turn to a discussion of physical implications of decay widths \eqref{decay_width_1}, \eqref{decay_width_2}, and \eqref{decay_width_3}. It is crucial to note that these decay widths share the following Planck mass suppression factor:
\begin{align}
  \left( G\,M^2 \right)^3 = \left(\cfrac{M}{M_\text{Pl}}
  \right)^6,
\end{align}
where $M_\text{Pl} = 1/\sqrt{G}$ is the Planck mass. Its present provides an extremely strong suppression. It is more convenient to operate with decay widths normalized by this factor:
\begin{align}\label{decay_widths}
  \begin{split}
    \overline{\Gamma}_{\text{S}\to\text{ss}} &\overset{\text{def}}{=} \cfrac{\Gamma_{\text{SS}\to\text{ss}}}{\left(G\,M^2\right)^3} = M \,  \cfrac{16}{3}\,\pi^6\, \, \sqrt{1-4\left(\cfrac{m}{M}\right)^2} \left[\mathcal{F}_1+2\,(1+\mathcal{F}_1) \left(\cfrac{m}{M}\right)^2\right]^2 \,, \\
    \overline{\Gamma}_{\text{S}\to\text{f}\overline{\text{f}}} & \overset{\text{def}}{=} \cfrac{ \Gamma_{\text{S}\to\text{f}\overline{\text{f}}} }{ \left(G\,M^2\right)^3 } = 384\,\pi^6 \, M \,\left(\cfrac{m_\text{f}}{M}\right)^2 \,\left[1 - 4 \left(\cfrac{m_\text{f}}{M}\right)^2\right]^{\frac32} \, \left(1+\mathcal{F}_2\right)^2\,,\\
    \overline{\Gamma}_{\text{S}\to\text{v}\overline{\text{v}}} &\overset{\text{def}}{=} \cfrac{\Gamma_{\text{S}\to\text{v}\overline{\text{v}}}}{ \left(G\,M^2\right)^3 } =  \cfrac{32}{3}\,\pi^6\,M\,.
  \end{split}
\end{align}
Their structure allows one to draw the following conclusions.

First and foremost, the decay to massless vectors is free from external parameters and the corresponding characteristic lifetime can be evaluated explicitly:
\begin{align}\label{lifetime_3}
  \tau_{\text{S}\to\text{v}\overline{\text{v}}} = \left(\cfrac{M_\text{Pl}}{M} \right)^7 \, \cfrac{3}{32 \,\pi^6} \,\cfrac{1}{M_\text{Pl}} \simeq 9.7515 \times 10^{-5} \left(\cfrac{M_\text{Pl}}{M}\right)^7 \, T_\text{Pl}.
\end{align}
Here $T_\text{Pl} = \hbar/( M_\text{Pl} \, c^2 ) \simeq 5.391 \times 10^{-44} \text{ sec}$ is the Planck time scale. The smaller the scalaron mass, the bigger the expected lifetime. For $M \sim 10^{-7} M_\text{Pl}$ the characteristic lifetime is about a second. For $M$ equal to the top quark mass the characteristic lifetime is $\tau_{\text{S}\to\text{v}\overline{\text{v}}} \sim 10^{64} \text{ sec}$ which is by $47$ order of magnitude more then the age of the Universe. Finally, the lifetime is equal to the age of the Universe for $M/M_\text{Pl} \sim 10^{-10}$ which marks the region of scalaron masses for which the discussed decay mechanism is relevant:
\begin{align}
  10^{-10} \ll \cfrac{M}{M_\text{Pl}} < 1 .
\end{align}

Secondly, for both scalar and fermion decay channels the results are inconclusive. The corresponding decay widths can be made arbitrary large in $\abs{\mathcal{F}_{1,2}} \to \infty$ limit. Moreover, for each channel it is possible to find critical values $\mathcal{F}_{1c}$, $\mathcal{F}_{2c}$ such that they make the corresponding decay widths vanish:
\begin{align}
  \mathcal{F}_{1c} &= - \cfrac{2 \left(\cfrac{m}{M}\right)^2}{ 1 + 2 \,\left(\cfrac{m}{M}\right)^2}\,, &\mathcal{F}_{2c} &= -1 \,.
\end{align}
Nonetheless, it is still possible to draw some meaningful conclusions. Namely, it is possible to maximize decay widths for the mass relation.

The fermion decay channel provides a simpler case. The decay width is maximal when the mass relation takes the following value:
\begin{align}
  \cfrac{m_\text{f}}{M} = \cfrac{1}{\sqrt{10}}\,.
\end{align}
This value does not depend on the renormalization constant $\mathcal{F}_2$. We set $\mathcal{F}_2=0$ because this allows us to operate with the UV-independent part of the amplitude. The corresponding maximal decay width and minimal characteristic lifetime read:
\begin{align}
  \overline{\Gamma}_{\text{S}\to\text{f}\overline{\text{f}},\text{max}} &= \cfrac{576}{25}\sqrt{\cfrac{3}{5}} \pi^6  M , & \tau_{\text{S}\to\text{f}\overline{\text{f}},\text{max}} = \left(\cfrac{M_\text{Pl}}{M}\right)^7 \cfrac{25}{576\,\pi^6} \sqrt{\cfrac{5}{3}} T_\text{Pl} \simeq 5.8283\times 10^{-5} \left(\cfrac{M_\text{Pl}}{M}\right)^7  T_\text{Pl}.
\end{align}
In full analogy with the previous case it is equal to the age of the Universe for $M/M_\text{Pl} \sim 10^{-10}$.

The scalar decay channel provides a more sophisticated case because the maximal value of the cross section is defined by the normalization constant $\mathcal{F}_1$:
\begin{align}\label{lifetime_2}
  \overline{\Gamma}_{\text{S}\to\text{ss},\text{max}} = \begin{cases}
    M \, \cfrac{64}{75}\,\pi^6 \left(1+3 \,\mathcal{F}_1\right)^2 \sqrt{\cfrac{1+3\, \mathcal{F}_1}{5+5\, \mathcal{F}_1}} & \mathcal{F}_1<-2 \text{ or } \mathcal{F}_1 > \xi \\
    M \, \cfrac{16}{3} \, \pi^6\,\mathcal{F}_1^2 & -2 < \mathcal{F}_1 < \xi
  \end{cases} \,.
\end{align}
Here $\mathcal{F}_1=-2$ and $\mathcal{F}_1 = \xi \simeq -0.14032$ are roots of the following equation:
\begin{align}
  \cfrac{4}{25} \,(1+3\,\mathcal{F}_1)^2 \,\sqrt{\cfrac{1+3\,\mathcal{F}_1}{5+5\,\mathcal{F}_1}} = \mathcal{F}_1^2 \,.
\end{align}
This expression of the maximal cross section is unbounded from above and reaches a minimum in point $\mathcal{F}_1 = \xi$. The corresponding characteristic lifetime reads:
\begin{align}\label{lifetime_1}
  \tau_{\text{S}\to\text{ss},\text{max}} = \left(\cfrac{M_\text{Pl}}{M}\right)^7 \, \cfrac{3}{16\,\pi^6}\, \cfrac{1}{\xi^2}\,T_\text{Pl} = 9.9052 \times 10^{-3} \left(\cfrac{M_\text{Pl}}{M}\right)^7 T_\text{Pl}.
\end{align}
The lifetime is equal to the age of the universe if $M/M_\text{Pl} \sim 10^{-9}$ which is pretty similar to previous cases. 

The results of this section shall be summarized as follows. First and foremost, decay width \eqref{decay_width_1} and \eqref{decay_width_2} strongly depends on the value of the renormalization constants $\mathcal{F}_1$ and $\mathcal{F}_2$. Despite this fact, it is possible to maximize the corresponding decay widths for the mass scale $M/M_\text{Pl}$. This allows one to establish meaningful lower bounds on the characteristic lifetimes \eqref{lifetime_1}, \eqref{lifetime_2}, \eqref{lifetime_3}. These results show that the presented decay mechanism is relevant only for the following range of mass hierarchy:
\begin{align}
  10^{-9} \ll \cfrac{M}{M_\text{Pl}} < 1\,.
\end{align}

\section{Discussion and conclusions}\label{section_conclusions}

In this paper we discussed new channels of reheating in the $R+R^2$ gravity model. The model has an additional scalar degree of freedom called scalaron. It drives the inflationary phase of expansions sliding the plain part of the potential. When the inflation ends with a graceful exit the Universe is filled with scalaron particles which shall reheat the Universe transferring their energy to the matter and dark matter sector.

We discuss new channels for such a reheating that take place because of the quantum gravitational effects. Perturbative quantum gravity provides a framework capable to account for such effects in a controllable manner. The first channel discussed in this paper is the annihilation of two scalaron in a pair of matter states. We obtained annihilation cross sections for massive scalars \eqref{cross_section_1}, massive Dirac fermions \eqref{cross_section_2}, and massless vectors \eqref{cross_section_3}. The second discussed channel is a decay of a single scalaron in a pair of two matter states that takes place at the one-loop level. Such processes are possible because the scalaron has a non-vanishing cubic self-coupling. It allows the scalaron to excite intermediate graviton states and to decay in pure matter states. We calculated decay widths for massive scalars \eqref{decay_width_1}, massive Dirac spinors \eqref{decay_width_2}, and massless vectors \eqref{decay_width_3}.

A detailed discussion of a reheating scenario that accounts for these new decay channels requires a separate treatment and lies far beyond the scope of this paper. However, it is possible to draw some meaningful conclusions about the influence of the presented processes.

Firstly, in the heavy scalaron limit production of fermions and massless vectors is strongly suppressed, so effectively only scalar particles are produced. At the same time, the discussed decay widths cannot be neglected in the heavy scalaron limit. Based on the reasoning provided in the previous section we can conclude that of scalaron mass $M \gg 10^{-9} M_\text{Pl}$ the characteristic scalaron lifetime is smaller than the age of the Universe. Therefore, the heavy scalaron limit contains a broad and interesting phenomenology as it can contribute to the reheating both through annihilation and decay channels.

Secondly, in the light scalaron limit the discussed scalaron decays are strongly suppressed by the factor $\left(M_\text{Pl}/M\right)^7$, which makes them negligibly small. The annihilation channel, on the contrary, provides a more sophisticated phenomenology. All types of particles are produced. Bosonic states (i.e. scalar states and vector states with a fixed chirality) are produced in equal amounts. The corresponding cross sections grow quadratic with the center-of-mass momentum. The production of fermions approaches a certain finite limiting value and growth no further. However, these cross sections do not have a direct influence on the production rate. The time evolution of the scalaron density is driven by the Zeldovich-Lee-Weinberg equation which contains an annihilation cross section after a certain thermodynamic averaging. We calculated the corresponding factors and established that fermion production is suppressed by the thermal factor $(T/M)$ while boson production experience a stronger thermal suppression $(T/M)^3$. Consequently, it is reasonable to assume that the boson production is dominant only at the early stage of reheating while the fermion production dominates the later stages.

The results of the present paper make it possible to study a reheating scenario in $R+R^2$ gravity which accounts for the discussed processes. Such research presents a separate sophisticated problem that will be discussed in further publications.

\section*{Acknowledgment}
The work was supported by the RSCF grant 22-22-00294. The author is grateful to A. Arbuzov, E. Arbuzova, and A. Dolgov for fruitful discussions.

\bibliographystyle{unsrturl}
\bibliography{SDiPQG.bib}

\begin{thebibliography}{10}

\bibitem{Accioly:2000nm}
A.~Accioly, S.~Ragusa, H.~Mukai, and E.~C. de~Rey~Neto.
\newblock {Algorithm for computing the propagator for higher derivative gravity
  theories}.
\newblock {\em Int. J. Theor. Phys.}, 39:1599--1608, 2000.
\newblock \href {https://doi.org/10.1023/A:1003632311419}
  {\path{doi:10.1023/A:1003632311419}}.

\bibitem{Hindawi:1995an}
Ahmed Hindawi, Burt~A. Ovrut, and Daniel Waldram.
\newblock {Consistent spin two coupling and quadratic gravitation}.
\newblock {\em Phys. Rev. D}, 53:5583--5596, 1996.
\newblock \href {http://arxiv.org/abs/hep-th/9509142}
  {\path{arXiv:hep-th/9509142}}, \href
  {https://doi.org/10.1103/PhysRevD.53.5583}
  {\path{doi:10.1103/PhysRevD.53.5583}}.

\bibitem{Dicke:1961gz}
R.~H. Dicke.
\newblock {Mach's principle and invariance under transformation of units}.
\newblock {\em Phys. Rev.}, 125:2163--2167, 1962.
\newblock \href {https://doi.org/10.1103/PhysRev.125.2163}
  {\path{doi:10.1103/PhysRev.125.2163}}.

\bibitem{Maeda:1988ab}
Kei-ichi Maeda.
\newblock {Towards the Einstein-Hilbert Action via Conformal Transformation}.
\newblock {\em Phys. Rev. D}, 39:3159, 1989.
\newblock \href {https://doi.org/10.1103/PhysRevD.39.3159}
  {\path{doi:10.1103/PhysRevD.39.3159}}.

\bibitem{Faraoni:1998qx}
Valerio Faraoni, Edgard Gunzig, and Pasquale Nardone.
\newblock {Conformal transformations in classical gravitational theories and in
  cosmology}.
\newblock {\em Fund. Cosmic Phys.}, 20:121, 1999.
\newblock \href {http://arxiv.org/abs/gr-qc/9811047}
  {\path{arXiv:gr-qc/9811047}}.

\bibitem{DeFelice:2010aj}
Antonio De~Felice and Shinji Tsujikawa.
\newblock {f(R) theories}.
\newblock {\em Living Rev. Rel.}, 13:3, 2010.
\newblock \href {http://arxiv.org/abs/1002.4928} {\path{arXiv:1002.4928}},
  \href {https://doi.org/10.12942/lrr-2010-3} {\path{doi:10.12942/lrr-2010-3}}.

\bibitem{Starobinsky:1980te}
Alexei~A. Starobinsky.
\newblock {A New Type of Isotropic Cosmological Models Without Singularity}.
\newblock {\em Phys. Lett. B}, 91:99--102, 1980.
\newblock \href {https://doi.org/10.1016/0370-2693(80)90670-X}
  {\path{doi:10.1016/0370-2693(80)90670-X}}.

\bibitem{Planck:2018jri}
Y.~Akrami et~al.
\newblock {Planck 2018 results. X. Constraints on inflation}.
\newblock {\em Astron. Astrophys.}, 641:A10, 2020.
\newblock \href {http://arxiv.org/abs/1807.06211} {\path{arXiv:1807.06211}},
  \href {https://doi.org/10.1051/0004-6361/201833887}
  {\path{doi:10.1051/0004-6361/201833887}}.

\bibitem{BICEP:2021xfz}
P.~A.~R. Ade et~al.
\newblock {Improved Constraints on Primordial Gravitational Waves using Planck,
  WMAP, and BICEP/Keck Observations through the 2018 Observing Season}.
\newblock {\em Phys. Rev. Lett.}, 127(15):151301, 2021.
\newblock \href {http://arxiv.org/abs/2110.00483} {\path{arXiv:2110.00483}},
  \href {https://doi.org/10.1103/PhysRevLett.127.151301}
  {\path{doi:10.1103/PhysRevLett.127.151301}}.

\bibitem{Paoletti:2022anb}
Daniela Paoletti, Fabio Finelli, Jussi Valiviita, and Masashi Hazumi.
\newblock {Planck and BICEP/Keck Array 2018 constraints on primordial
  gravitational waves and perspectives for future B-mode polarization
  measurements}.
\newblock 8 2022.
\newblock \href {http://arxiv.org/abs/2208.10482} {\path{arXiv:2208.10482}}.

\bibitem{Vilenkin:1985md}
Alexander Vilenkin.
\newblock {Classical and Quantum Cosmology of the Starobinsky Inflationary
  Model}.
\newblock {\em Phys. Rev. D}, 32:2511, 1985.
\newblock \href {https://doi.org/10.1103/PhysRevD.32.2511}
  {\path{doi:10.1103/PhysRevD.32.2511}}.

\bibitem{Koshelev:2016xqb}
Alexey~S. Koshelev, Leonardo Modesto, Leslaw Rachwal, and Alexei~A.
  Starobinsky.
\newblock {Occurrence of exact $R^2$ inflation in non-local UV-complete
  gravity}.
\newblock {\em JHEP}, 11:067, 2016.
\newblock \href {http://arxiv.org/abs/1604.03127} {\path{arXiv:1604.03127}},
  \href {https://doi.org/10.1007/JHEP11(2016)067}
  {\path{doi:10.1007/JHEP11(2016)067}}.

\bibitem{Arbuzova:2011fu}
E.~V. Arbuzova, A.~D. Dolgov, and L.~Reverberi.
\newblock {Cosmological evolution in $R^2$ gravity}.
\newblock {\em JCAP}, 02:049, 2012.
\newblock \href {http://arxiv.org/abs/1112.4995} {\path{arXiv:1112.4995}},
  \href {https://doi.org/10.1088/1475-7516/2012/02/049}
  {\path{doi:10.1088/1475-7516/2012/02/049}}.

\bibitem{Arbuzova:2018ydn}
E.~V. Arbuzova, A.~D. Dolgov, and R.~S. Singh.
\newblock {Distortion of the standard cosmology in $R+R^2$ theory}.
\newblock {\em JCAP}, 07:019, 2018.
\newblock \href {http://arxiv.org/abs/1803.01722} {\path{arXiv:1803.01722}},
  \href {https://doi.org/10.1088/1475-7516/2018/07/019}
  {\path{doi:10.1088/1475-7516/2018/07/019}}.

\bibitem{Arbuzova:2021etq}
Elena Arbuzova, Alexander Dolgov, and Rajnish Singh.
\newblock {$R^2$-Cosmology and New Windows for Superheavy Dark Matter}.
\newblock {\em Symmetry}, 13(5):877, 2021.
\newblock \href {https://doi.org/10.3390/sym13050877}
  {\path{doi:10.3390/sym13050877}}.

\bibitem{Latosh:2020ysu}
B.~N. Latosh.
\newblock {Basic Problems of Conservative Approaches to a Theory of Quantum
  Gravity}.
\newblock {\em Phys. Part. Nucl.}, 51(5):859--878, 2020.
\newblock \href {http://arxiv.org/abs/2003.02462} {\path{arXiv:2003.02462}},
  \href {https://doi.org/10.1134/S1063779620050056}
  {\path{doi:10.1134/S1063779620050056}}.

\bibitem{Burgess:2003jk}
C.~P. Burgess.
\newblock {Quantum gravity in everyday life: General relativity as an effective
  field theory}.
\newblock {\em Living Rev. Rel.}, 7:5--56, 2004.
\newblock \href {http://arxiv.org/abs/gr-qc/0311082}
  {\path{arXiv:gr-qc/0311082}}, \href {https://doi.org/10.12942/lrr-2004-5}
  {\path{doi:10.12942/lrr-2004-5}}.

\bibitem{Levi:2018nxp}
Mich\`ele Levi.
\newblock {Effective Field Theories of Post-Newtonian Gravity: A comprehensive
  review}.
\newblock {\em Rept. Prog. Phys.}, 83(7):075901, 2020.
\newblock \href {http://arxiv.org/abs/1807.01699} {\path{arXiv:1807.01699}},
  \href {https://doi.org/10.1088/1361-6633/ab12bc}
  {\path{doi:10.1088/1361-6633/ab12bc}}.

\bibitem{Calmet:2013hfa}
Xavier Calmet.
\newblock {Effective theory for quantum gravity}.
\newblock {\em Int. J. Mod. Phys. D}, 22:1342014, 2013.
\newblock \href {http://arxiv.org/abs/1308.6155} {\path{arXiv:1308.6155}},
  \href {https://doi.org/10.1142/S0218271813420145}
  {\path{doi:10.1142/S0218271813420145}}.

\bibitem{Vanhove:2021zel}
Pierre Vanhove.
\newblock {$S$-matrix approach to general gravity and beyond}.
\newblock In {\em {55th Rencontres de Moriond on QCD and High Energy
  Interactions}}, 4 2021.
\newblock \href {http://arxiv.org/abs/2104.10148} {\path{arXiv:2104.10148}}.

\bibitem{tHooft:1974toh}
Gerard 't~Hooft and M.~J.~G. Veltman.
\newblock {One loop divergencies in the theory of gravitation}.
\newblock {\em Ann. Inst. H. Poincare Phys. Theor. A}, 20:69--94, 1974.

\bibitem{Goroff:1985sz}
Marc~H. Goroff and Augusto Sagnotti.
\newblock {Quantum Gravity at Two Loops}.
\newblock {\em Phys. Lett. B}, 160:81--86, 1985.
\newblock \href {https://doi.org/10.1016/0370-2693(85)91470-4}
  {\path{doi:10.1016/0370-2693(85)91470-4}}.

\bibitem{Prinz:2020nru}
David Prinz.
\newblock {Gravity-Matter Feynman Rules for any Valence}.
\newblock {\em Class. Quant. Grav.}, 38(21):215003, 2021.
\newblock \href {http://arxiv.org/abs/2004.09543} {\path{arXiv:2004.09543}},
  \href {https://doi.org/10.1088/1361-6382/ac1cc9}
  {\path{doi:10.1088/1361-6382/ac1cc9}}.

\bibitem{DeWitt:1967uc}
Bryce~S. DeWitt.
\newblock {Quantum Theory of Gravity. 3. Applications of the Covariant Theory}.
\newblock {\em Phys. Rev.}, 162:1239--1256, 1967.
\newblock \href {https://doi.org/10.1103/PhysRev.162.1239}
  {\path{doi:10.1103/PhysRev.162.1239}}.

\bibitem{Sannan:1986tz}
Sigurd Sannan.
\newblock {Gravity as the Limit of the Type {II} Superstring Theory}.
\newblock {\em Phys. Rev. D}, 34:1749, 1986.
\newblock \href {https://doi.org/10.1103/PhysRevD.34.1749}
  {\path{doi:10.1103/PhysRevD.34.1749}}.

\bibitem{Latosh:2022ydd}
Boris Latosh.
\newblock {FeynGrav : FeynCalc extension for gravity amplitudes}.
\newblock {\em Class. Quant. Grav.}, 1 2022.
\newblock \href {http://arxiv.org/abs/2201.06812} {\path{arXiv:2201.06812}},
  \href {https://doi.org/10.1088/1361-6382/ac7e15}
  {\path{doi:10.1088/1361-6382/ac7e15}}.

\bibitem{Mertig:1990an}
R.~Mertig, M.~Bohm, and Ansgar Denner.
\newblock {FEYN CALC: Computer algebraic calculation of Feynman amplitudes}.
\newblock {\em Comput. Phys. Commun.}, 64:345--359, 1991.
\newblock \href {https://doi.org/10.1016/0010-4655(91)90130-D}
  {\path{doi:10.1016/0010-4655(91)90130-D}}.

\bibitem{Shtabovenko:2020gxv}
Vladyslav Shtabovenko, Rolf Mertig, and Frederik Orellana.
\newblock {FeynCalc 9.3: New features and improvements}.
\newblock {\em Comput. Phys. Commun.}, 256:107478, 2020.
\newblock \href {http://arxiv.org/abs/2001.04407} {\path{arXiv:2001.04407}},
  \href {https://doi.org/10.1016/j.cpc.2020.107478}
  {\path{doi:10.1016/j.cpc.2020.107478}}.

\bibitem{Patel:2015tea}
Hiren~H. Patel.
\newblock {Package-X: A Mathematica package for the analytic calculation of
  one-loop integrals}.
\newblock {\em Comput. Phys. Commun.}, 197:276--290, 2015.
\newblock \href {http://arxiv.org/abs/1503.01469} {\path{arXiv:1503.01469}},
  \href {https://doi.org/10.1016/j.cpc.2015.08.017}
  {\path{doi:10.1016/j.cpc.2015.08.017}}.

\bibitem{Patel:2016fam}
Hiren~H. Patel.
\newblock {Package-X 2.0: A Mathematica package for the analytic calculation of
  one-loop integrals}.
\newblock {\em Comput. Phys. Commun.}, 218:66--70, 2017.
\newblock \href {http://arxiv.org/abs/1612.00009} {\path{arXiv:1612.00009}},
  \href {https://doi.org/10.1016/j.cpc.2017.04.015}
  {\path{doi:10.1016/j.cpc.2017.04.015}}.

\bibitem{Shtabovenko:2016whf}
Vladyslav Shtabovenko.
\newblock {FeynHelpers: Connecting FeynCalc to FIRE and Package-X}.
\newblock {\em Comput. Phys. Commun.}, 218:48--65, 2017.
\newblock \href {http://arxiv.org/abs/1611.06793} {\path{arXiv:1611.06793}},
  \href {https://doi.org/10.1016/j.cpc.2017.04.014}
  {\path{doi:10.1016/j.cpc.2017.04.014}}.

\bibitem{Mandelstam:1959bc}
Stanley Mandelstam.
\newblock {Analytic properties of transition amplitudes in perturbation
  theory}.
\newblock {\em Phys. Rev.}, 115:1741--1751, 1959.
\newblock \href {https://doi.org/10.1103/PhysRev.115.1741}
  {\path{doi:10.1103/PhysRev.115.1741}}.

\bibitem{Mandelstam:1958xc}
S.~Mandelstam.
\newblock {Determination of the pion - nucleon scattering amplitude from
  dispersion relations and unitarity. General theory}.
\newblock {\em Phys. Rev.}, 112:1344--1360, 1958.
\newblock \href {https://doi.org/10.1103/PhysRev.112.1344}
  {\path{doi:10.1103/PhysRev.112.1344}}.

\bibitem{Bilenky:1995zq}
Samoil~M. Bilenky.
\newblock {\em {Introduction to Feynman diagrams and electroweak interactions
  physics}}.
\newblock 1995.

\bibitem{Weinberg:1995mt}
Steven Weinberg.
\newblock {\em {The Quantum theory of fields. Vol. 1: Foundations}}.
\newblock Cambridge University Press, 6 2005.

\bibitem{Peskin:1995ev}
Michael~E. Peskin and Daniel~V. Schroeder.
\newblock {\em {An Introduction to quantum field theory}}.
\newblock Addison-Wesley, Reading, USA, 1995.

\bibitem{Zeldovich:1965gev}
Ya.~b. Zeldovich.
\newblock {Survey of Modern Cosmology}.
\newblock {\em Adv. Astron. Astrophys.}, 3:241--379, 1965.
\newblock \href {https://doi.org/10.1016/b978-1-4831-9921-4.50011-9}
  {\path{doi:10.1016/b978-1-4831-9921-4.50011-9}}.

\bibitem{Lee:1977ua}
Benjamin~W. Lee and Steven Weinberg.
\newblock {Cosmological Lower Bound on Heavy Neutrino Masses}.
\newblock {\em Phys. Rev. Lett.}, 39:165--168, 1977.
\newblock \href {https://doi.org/10.1103/PhysRevLett.39.165}
  {\path{doi:10.1103/PhysRevLett.39.165}}.

\bibitem{Passarino:1978jh}
G.~Passarino and M.~J.~G. Veltman.
\newblock {One Loop Corrections for e+ e- Annihilation Into mu+ mu- in the
  Weinberg Model}.
\newblock {\em Nucl. Phys. B}, 160:151--207, 1979.
\newblock \href {https://doi.org/10.1016/0550-3213(79)90234-7}
  {\path{doi:10.1016/0550-3213(79)90234-7}}.

\end{thebibliography}

\end{document}